\documentclass[12pt]{article}
\usepackage{amssymb}
\begin{document} 

\begin{center}
{\bf \Large Multiple Landen values and the tribonacci numbers}\\[10pt]
{\bf \large David Broadhurst}\\[5pt]
{Department of Physical Sciences, Open University, Milton Keynes MK7 6AA, UK\\
Institut f\"{u}r Mathematik und Institut f\"{u}r Physik, Humboldt-Universit\"{a}t zu Berlin}\\[5pt]
24 April 2015
\end{center}

{\large
Multiple Landen values (MLVs) are defined as iterated integrals
on the interval $x\in[0,1]$ of the differential forms
$A=d\log(x)$, $B=-d\log(1-x)$, $F=-d\log(1-\rho^2x)$
and $G=-d\log(1-\rho x)$, where $\rho=(\sqrt{5}-1)/2$
is the golden section.
I conjecture that the dimension
of the space of ${\mathbb Z}$-linearly independent
MLVs of weight $w$ is a tribonacci number $T_w$, generated by
$1/(1-x-x^2-x^3)=1+\sum_{w>0}T_w x^w$, and that a basis
is provided by all the words in the $\{A,G\}$ sub-alphabet that
neither end in $A$ nor contain $A^3$.
For $w<9$, I construct a much more efficient basis,
for a MLV datamine, where no prime greater
than 11 occurs in the denominators of 3,357,257
coefficients of rational reduction of 49,151 MLVs. Numerical
data for 40 primitives then enable fast evaluation of all
of these MLVs to 20,000 digits. The datamine provides reductions
of Ap\'ery-type sums
$A_w=\sum_{n>0}(-1)^{n+1}n^{-w}/{2n\choose n}$
and 6 ladder-combinations of depth-1 polylogarithms
${\rm Li}_w(\rho^p)=\sum_{n>0}\rho^{pn}n^{-w}$
with $p\in\{1,2,3,4,6,8,10,12,20,24\}$ and
coefficients  given by Landen, Coxeter and Lewin at $w=2$.
I prove that the former  evaluate to MLVs and conjecture that  the
latter do. Comparison is made between the properties of MLVs
and multiple polylogarithms at roots of unity, encountered
in the quantum field theory of the standard model of particle physics.

\newpage
\section{Introduction}

In 1780, John Landen, a land-agent~\cite{DNB,Watson}
in the English county of Northamptonshire, adjacent to my own county of residence,
published a book,  {\em Mathematical memoirs respecting a
variety of subjects}~\cite{L}. Memoir V,
{\em A new method of obtaining the sums of certain series}, gives
reductions of the dilogarithm ${\rm Li}_2(x)=\sum_{n>0}x^n/n^2$
to rational combinations of $\pi^2$ and squares of logarithms, for the special
values $x=\frac12$, $x=\rho$ and $x=\rho^2=1-\rho$, where $\rho=(\sqrt{5}-1)/2$
is the {\em golden section}. The case $x=\frac12$ had been studied
by Euler and independently by Landen, 20 years earlier~\cite{L6},
but Landen's golden results were new.
In the subsequent 235 years, no further result of this type has been found, for $1>x>0$.
Landen also reduced the trilogarithmic combinations
${\rm Li}_3(\frac12)-\frac78{\rm Li}_3(1)$ and ${\rm Li}_3(\rho^2)-\frac45{\rm Li}_3(1)$
to products of polylogarithms of lesser weight.
No other such relation between a pair of polylogarithms has been found~\cite{Lewin91}.

To achieve these feats,  Landen exploited the happy fact that polylogarithms are iterated fluents,
or iterated
integrals, as we now call them. Thus, for example, he proved that
$$x+\frac{x^2}{2^2}+\frac{x^3}{3^2}+\frac{x^4}{4^2}~\&{\rm c}.=
{\rm fl}.~\overline{\frac{\dot{x}}{x}\cdot{\rm fl}.~\frac{\dot{x}}{1-x}}
=\frac25a^2-{\rm sq}.~{\rm Log}.~x$$
for $x=(\sqrt{5}-1)/2$ and $a$ defined as the length of an arc of a quadrant of a circle of
unit radius. In modern terms, that evaluates
\begin{equation}
{\rm Li}_2(\rho)\equiv\sum_{n=1}^\infty\frac{\rho^n}{n^2}=
\int_0^\rho\frac{dx_1}{x_1}\int_0^{x_1}\frac{dx_2}{1-x_2}=\frac25\left(\frac{\pi}{2}\right)^2-(\log(\rho))^2.
\label{Landen}
\end{equation}

The present work is devoted to the study of multiple {\em Landen} values (MLVs), which I
define as iterated integrals,
on the interval $x\in[0,1]$, of the differential
forms $A=d\log(x)$, $B=-d\log(1-x)$, $F=-d\log(1-\rho^2x)$
and the {\em golden} letter  $G=-d\log(1-\rho x)$. By $Z(W)$, I denote the map from a word $W$
in the $\{A,B,F,G\}$ alphabet to the iterated integral encoded by the letters of $W$.
The {\em weight} of a word $W$ is the number of letters in $W$ and its {\em depth}
is the number of letters not equal to $A$. Thus Landen's result~(\ref{Landen})
may be re-written as $Z(AG)=\frac35Z(AB)-\frac12Z(GG)$, which yields the integer relation
$6Z(AB)=10Z(AG)+5Z(GG)$, at weight $w=2$. At $w=3$, the integer relations
$2Z(ABB)=2Z(AAB)=10Z(AGG)+5Z(GGG)$ likewise reduce multiple {\em zeta} values  (MZVs),
formed from words in the sub-alphabet $\{A,B\}$, to multiple {\em golden}  values (MGVs),
formed from words in the golden sub-alphabet $\{A,G\}$.

Theorem~1 shows that something
quite new happens at $w=4$, where there is an integer relation between MGVs.
This is the beginning
of a wonderful sequence of relations between MGVs which lead me to claim in Conjecture~1
that the dimension of the space of ${\mathbb Z}$-linearly independent
finite MGVs of weight $w$ is a tribonacci number $T_w$, generated by
$1/(1-x-x^2-x^3)=1+\sum_{w>0}T_w x^w$. Conjecture~2 is even bolder and claims that
$T_w$ is also the dimension of the space of ${\mathbb Z}$-linearly independent
finite MLVs of weight $w$ in the {\em full} alphabet $\{A,B,F,G\}$. That implies the existence
of precisely 36,783 integer relations between MLVs at $w=8$, all of which
are now recorded in a datamine of MLVs, available on request to the author.

In the course of this work,
I obtain Theorem~2, on the alternating binomial sums~\cite{A}
$A_w=\sum_{n>0}(-1)^{n+1}n^{-w}/{2n\choose n}$
and give a novel conjecture for
6 ladder-combinations of depth-1 polylogarithms
${\rm Li}_w(\rho^p)=\sum_{n>0}\rho^{pn}n^{-w}$
with exponents $p\in\{1,2,3,4,6,8,10,12,20,24\}$ and
coefficients  given by Landen~\cite{L}, Coxeter~\cite{Cox} and
Lewin~\cite{Lewin85,Lewin82,Lewin84,Lewin86,Lewin91} at $w=2$.

 This paper is organized as follows. Section~2 explains
 a crucial difference between MZVs and MLVs.
 Multiple zeta values enjoy two algebras, coming from the shuffles
 of iterated integrals and the stuffles of nested sums, but for multiple
 Landen values the algebra of nested sums is not closed. This makes
 it rather hard to prove many of the fascinating structural relations
 between MLVs discovered in the course of my work. Section~3
 gives a modest number of proofs; Section~4 contains a wealth
 of empirical results and stringently investigated conjectures.
 In Section~5, I turn attention to the ladder relations of~\cite{Lewin91},
 which end at $w=9$. Yet I conjecture that  the 6 combinations
of polylogarithms to which these relations refer are always reducible to MGVs.
Section~6 compares and contrasts MLVs with iterated integrals encountered
in my chosen specialism of quantum field theory, whose practical agenda
has done much to enrich the study of iterated integrals. Section~7 offers conclusions.

\section{Many shuffles but fewer stuffles}

The product $Z(U)Z(V)$ of a pair of MLVs is a sum of MLVs, namely $\sum_{W\in{\cal S}(U,V)}Z(W)$
where ${\cal S}(U, V )$ is the set of all words $W$ that result from shuffling the words $U$ and $V$ .
Shuffles preserve the order of letters in $U$ and the order of letters in $V$, but are otherwise
unconstrained. Thus, for example, we obtain the shuffle product
$$Z(AB)Z(GF) = Z(A(BGF+GBF+GFB)+G(ABF+AFB+FAB))$$
with a notation in which $\sum_n Z(c_n W_n)=\sum_n c_n Z(W_n)$, for real  $c_n$. Note that
each term on the right has weight $w=4$ and depth $d=3$, which are the sums
of the weights and depths of $Z(AB)$ and $Z(GF)$.

If $W$ is a word of weight $w$ and depth $d$ in the alphabet $\{A,B,F,G\}$ and
$W$ neither begins with $B$ nor ends with $A$, then $Z(W)$ is a finite MLV that may be written
as a $d$-fold sum of the form
\begin{equation}
{\rm Li}_{a_1,a_2,\ldots,a_d}(z_1,z_2,\ldots,z_d)\equiv\sum_{n_1>n_2>\ldots>n_d>0}
\quad\prod_{j=1}^d\frac{z_j^{n_j}}{n_j^{a_j}}.
\label{Li}
\end{equation}
To determine the arguments, let $L_j$ be the $j$-th letter in $W$ that is not $A$.
Then $a_j-1$ is the exponent of $A$ before $L_j$. If $L_j=B$, set $p_j=0$;
if  $L_j=F$, set $p_j=2$; if $L_j=G$, set $p_j=1$. Then the exponent of $\rho$ in $z_j$
is  $p_j-p_{j-1}$, with  $p_0=0$. Thus $Z(AAABAGFAAB)={\rm Li}_{4,2,1,3}(1,\rho,\rho,1/\rho^2)$.

The nested sums~(\ref{Li}) are endowed with a stuffle algebra that preserves weight but not depth.
For example, the stuffle product ${\rm Li}_a(x){\rm Li}_b(y)=
{\rm Li}_{a,b}(x,y)+{\rm Li}_{b,a}(y,x)+{\rm Li}_{a+b}(xy)$
contains a depth-1 term, coming from coincidence of indices of summation.
Unfortunately this is often useless for constraining MLVs. None of the three terms in
$Z(AF)Z(AG)={\rm Li}_{2,2}(\rho^2,\rho)+{\rm Li}_{2,2}(\rho,\rho^2)+{\rm Li}_4(\rho^3)$ is a MLV.
The stuffle product of a MZV and a  MLV  gives MLVs, as does the stuffle product
of a MGV with a MGV. Thus the stuffle relations
\begin{eqnarray*}
Z(AB)Z(AF)&=&Z(ABAF+AFAF)+Z(AAAF)\\
Z(AG)Z(GG)&=&Z(AGFF+GAFF+GGAF)+Z(AAFF+GAAF)
\end{eqnarray*}
add new information to the depth-conserving shuffles.
However, none of the information gained from the many shuffles and the fewer stuffles
uses the golden relation $\rho^2=1-\rho$, to which I now turn.

\section{A modicum of proof}

I present a slender, yet seminal, body of proof,
before presenting empirical findings.
A pragmatic reader may skip to Subsection~3.4.

\subsection{An integer relation at weight 4}

Let $W$ be a word in a binary alphabet $\{A,B\}$  with
$A=d\log(x)$ and $B=-d\log(1-x)$. If $W$ does not end in $A$,
let $L(W,y)$ be the iterated integral from $x=0$ to $x=y$
of the sequence of differential forms encoded by $W$.
Hence $y L^\prime(AW,y)=(1-y)L^\prime(BW,y)=L(W,y)$.
We declare that $L(1,y)=1$, with unity denoting the empty word,
and linearly extend by $\sum_n L(c_n W_n,y)=\sum_n c_n L(W_n,y)$, for real $c_n$.
Then $L(W(A,B),y)=Z(W(A,Y))$ with $Y=-d\log(1-xy)$ replacing $B=-d\log(1-x)$.

For any $y\in[0,1]$ and $W$ not beginning with $B$,  we have a MZV evaluation
\begin{equation}
Z(W)\equiv L(W,1)=\sum_{W=UV}L(\widetilde{U},1-y)L(V,y)=Z(\widetilde{W})
\label{tilde}
\end{equation}
where the sum is over all deconcatenations~\cite{trans} of $W$ into
a first part, $U$, and a second part, $V$, and $\widetilde{U}$
is the dual of $U$, obtained by reversing the order of letters and
exchanging $A$ and $B$. Thus the dual of $AAB$ is $ABB$.
For a MZV of weight $w$ there are $w+1$ deconcatenations in~(\ref{tilde}),
corresponding to the places that $y$ may sit inside the inequalities
$1>x_1>\ldots>x_w>0$ for the integration variables.
The integrations with $y>x_i$ yield $L(V,y)$ and those with $x_j>y$ yield
$L(\widetilde{U},1-y)$, after transforming $x_j\to1-x_j$.

Now let us, {\em pro tempore}, discard products, denoting their neglect by $\simeq$. Then
by setting $y=\rho^2$ in~(\ref{tilde}) we obtain
\begin{equation}
Z(W)\simeq L(\widetilde{W},\rho)+L(W,\rho^2)
\label{sim}
\end{equation}
where the first term is a MGV in the $\{A,G\}$  sub-alphabet
and the second is a MLV in the $\{A,F\}$ sub-alphabet.

{\bf Lemma 1:} Every MLV in the $\{A,F\}$ sub-alphabet is a ${\mathbb Q}$-linear
combination of MZVs, MGVs and products
of these two types of term.

{\bf Proof:} For weight $w>1$, suppose that this is true for smaller weights.
Let $W$ be a Lyndon word in the $\{A,B\}$  sub-alphabet, namely
a word for which all deconcatenations $W=UV$ have $U$
preceding $V$, in lexicographic ordering. Then we may use~(\ref{sim}) at weight $w$,
since omitted products are, by assumption, of the required form.
Hence the MLV obtained by replacing
$B$ by $F$ in the Lyndon word $W$ is also of the required form. Thus, at weight $w$, all MLVs in the
$\{A,F\}$ sub-alphabet are of the required form, since the shuffle algebra
gives them in terms of Lyndon words and products.
The observation that $2Z(F)=Z(G)$ completes the proof by induction.~$\blacksquare$

Relation~(\ref{tilde}) acquires more power when we combine it with
\begin{equation}
L(W,-\rho)=L(\overline{W},\rho^2)
\label{barr}
\end{equation}
with a conjugate defined by  $\overline W(A,B)\equiv W(A+B,-B)$.
The operations of taking a conjugate (denoted by bar) or a dual (denoted by tilde)
are self-inverse and commute. Thus if
$W=\overline{U}=\widetilde{V}$, then $U=\overline{W}$, $V=\widetilde{W}$ and
$\widetilde{U}=\overline{V}$.
Relation~(\ref{barr}) results from  transformation of variables of integration by $x_j\to -x_j/(1-x_j)$,
combined with the identity $\rho/(1+\rho)=\rho^2$.

Next consider the depth-1 word $W=A^{w-1}B$, which yields the
polylogarithm $L(A^{w-1}B,y)={\rm Li}_w(y)$
for which the transformation of integration variables
$x_j\to x_j^2$ gives
${\rm Li}_w(y)=2^{1-w}{\rm Li}(y^2)-{\rm Li}_w(-y)$.
Setting $y=\rho$ and using~(\ref{barr}), we obtain
\begin{equation}
Z(A^{w-1}G)=2^{1-w}Z(A^{w-1}F)+Z((A+F)^{w-1}F)
\label{dbl}
\end{equation}
with a final term that yields $2^{w-1}$ distinct MLVs, with depths
up to $w$, when we expand $(A+F)^{w-1}$, remembering that
$A$ and $F$ do not commute.

Then we expect to be able
to reduce all MZVs with weight $w<8$ to MGVs,
by diligent use of~(\ref{tilde}) and (\ref{dbl}), and hence
by Lemma~1 to reduce all MLVs with $w<8$ in the
$\{A,F\}$ sub-alphabet to words in the golden sub-alphabet.
To show that we may do so, it is necessary to prove
that $\zeta(2)$, $\zeta(3)$, $\zeta(5)$ and $\zeta(7)$ are reducible to MGVs,
which I shall do.

At $w=2,3$, we obtain intriguingly similar reductions of MZVs to MGVs
\begin{eqnarray}
6\zeta(2)&=&5Z((2A+G)G)\label{z2}\\
2\zeta(3)&=&5Z((2A+G)G^2)\label{z3}
\end{eqnarray}
with~(\ref{z2}) being Landen's result $\pi^2=10\,{\rm Li}_2(\rho)+10(\log(\rho))^2$,
obtained by simple algebra as follows. First use~(\ref{dbl}) to show that $Z(G)=2Z(F)$
and that $Z(AG)=\frac{3}{2}Z(AF)+Z(FF)$.
The shuffle algebra gives $Z(G^w)=(Z(G))^w/w!=Z(F^w)/2^w$.
Hence we reduce $Z(AF)=\frac{2}{3}Z(AG)-\frac16Z(GG)$ to
MGVs. Then we obtain $\zeta(2)=Z(AF)+Z(F)Z(G)+Z(AG)$
from~(\ref{tilde}) and lift the product $Z(F)Z(G)$ to $Z(GG)$,
obtaining~(\ref{z2}). Similar (but more tedious) elimination and lifting
at $w=3$ give the neat result~(\ref{z3}), from which the depth-1 sum
$Z(AAG)={\rm Li}_3(\rho)$
is notably absent.

We now know, from Lemma~1, that all MLVs in the
$\{A,F\}$ alphabet with $w<5$ are reducible to MGVs, since there is no primitive MZV
at $w=4$.

{\bf Theorem 1:} There is an integer relation between MGVs of weight 4.

{\bf Proof:} To prove existence we do not need to retain products.
From~(\ref{dbl}) we obtain $Z(AAAG)\simeq\frac98Z(AAAF)+Z(U)+Z(V)$
where $U=FAAF+AFAF+AAFF$ and $V=FFAF+FAFF+AFFF$. The shuffle products
$Z(F)Z(AAF)=Z(FAAF+AFAF+2AAFF)$ and $Z(F)Z(FAF)=2Z(FFAF+FAFF)$
give $Z(U)\simeq -Z(AAFF)$ and $Z(V)\simeq Z(AFFF)$. Next
we obtain $Z(AFFF+AAAG)\simeq Z(AAFF+AAGG)\simeq Z(AAAF+AGGG)
\simeq0$ from~(\ref{sim}), since all MZVs of weight 4 are rational multiples
of $\pi^4$ and hence products of MGVs. Collecting terms we obtain
\begin{equation}
8Z(A^2G^2)\simeq16Z(A^3G)+9Z(AG^3)
\label{proof1}
\end{equation}
with neglect of products that may be lifted to MGVs.
This lifting cannot trivialize~(\ref{proof1}), because it contains
only Lyndon words.~$\blacksquare$

After some rather heavy lifting duty, we obtain the relation explicitly as
\begin{equation}
48Z((2A+G)AAG)=Z((2A+G)(176AG+76GA+123GG)G)
\label{id4}
\end{equation}
with less attractive integers than in the Lyndonized version~(\ref{proof1}),
but now a pleasing ubiquity of $(2A+G)$.

\subsection{A theorem for Ap\'ery's alternating binomial sums}

Now consider the Ap\'ery-type alternating binomial sums~\cite{A,BBK}
\begin{equation}
A_w\equiv\sum_{n>0}\frac{(-1)^{n+1}}{n^w{2n\choose n }}.
\label{Aw}\end{equation}

{\bf Theorem 2:} For $w>1$,  $A_w$ is a ${\mathbb Z}$-linear combination of MGVs.

{\bf Proof:} Multiply the summand in~(\ref{Aw}) by
$$2n^w\int_0^1\left(\frac{(-2\log(y))^{w-1}}{(w-1)!}\right)y^{2n-1}dy=1$$
and exchange the order of summation and integration to obtain
$$A_w=\int_0^1\left(\frac{(-2\log(y))^{w-1}}{(w-1)!}\right)g(y)dy$$
where $g(y)=f^\prime(y)$ is the derivative of the 
summable series~\cite{BBK,binom}
$$f(y)\equiv-\sum_{n>0}\frac{(-y^2)^n}{n{2n\choose n}}=
\frac{y\log\left(\sqrt{1+y^2/4}+y/2\right)}{\sqrt{1+y^2/4}}.$$
Hence $A_1=f(1)=-2\log(\rho)/\sqrt{5}=Z(G)/\sqrt{5}$ and for $w>1$ we obtain
$$A_w=2\int_0^1\left(\frac{(-2\log(y))^{w-2}}{(w-2)!}\right)
\frac{\log\left(\sqrt{1+y^2/4}+y/2\right)}{\sqrt{1+y^2/4}}dy$$
using integration by parts. Now make the substitution $y=\rho x/\sqrt{1-\rho x}$,
which maps $y=1$ to $x=1$ and gives $A_w=I_{w-2,2}$, where
\begin{equation}
I_{a,b}\equiv\int_0^1\frac{(\log(1-\rho x)-2\log(\rho)-2\log(x))^a(-\log(1-\rho x))^{b-1}\rho\,dx}
{a!(b-1)!(1-\rho x)}
\label{iab}
\end{equation}
and a trinomial expansion gives
\begin{equation}
I_{a,b}=\sum_{i+j+k=a}(-1)^k{k+b-1\choose b-1}Z(G^i)Z((2A)^jG^{k+b}).
\label{trin}
\end{equation}
Then the shuffle algebra lifts this to a ${\mathbb Z}$-linear combination of  MGVs .~$\blacksquare$

The lift of~(\ref{trin}) is $I_{a,b}=Z((2A+G)^aG^b)$, which delivers
$A_2=I_{0,2}=Z(G^2)=2(\log(\rho))^2$ and, on use of~(\ref{z3}),
$A_3=I_{1,2}=Z((2A+G)G^2)=\frac25\zeta(3)$, from which Ap\'ery~\cite{A},
with great ingenuity~\cite{missed}, proved the irrationality of $\zeta(3)$.
Thereafter, expansion~(\ref{trin})  is more revealing, since it tells us the primitive part
of $A_w$ directly, as a sum of $(w-2)$ Lyndon words:
\begin{equation}
A_w=Z((2A+G)^{w-2}G^2)\simeq\sum_{k=0}^{w-3}(-1)^k(k+1)Z((2A)^{w-2-k}G^{k+2}).
\label{trins}
\end{equation}
Hence $A_4\simeq 4Z(A^2G^2-AG^3)$, from which we eliminate the depth-2 term,
using~(\ref{proof1}), and transform the depth-3 term,  using $Z(AG^3+A^3F)\simeq0$,
to obtain the primitive part of
\begin{equation}
A_4=Z((2A+G)^2G^2)\simeq\frac12 Z(A^3(16G-F))=\sum_{n>0}\frac{8\rho^{2n-1}}{(2n-1)^4}
\label{a4}
\end{equation}
as a depth-1 sum over {\em odd} powers of the golden section.

I now prove a result conjectured in~\cite{BBK},
namely that the primitive part of
\begin{equation}
A_5=Z((2A+G)^3G^2)\simeq\frac12Z(A^4(5F-4B))=\sum_{n>0}\frac{5\rho^{2n}-4}{2n^5}
\label{a5}
\end{equation}
is a depth-1 sum with only {\em even} powers of $\rho$ appearing.
To do this, use the shuffle algebra to remove products from~(\ref{dbl}),
in the manner that we did in~(\ref{trins}), obtaining
\begin{equation}
Z(A^{w-1}(G-(2^{1-w}+1)F))\simeq F_w\equiv\sum_{k=1}^{w-2}(-1)^k Z(A^{w-1-k}F^{k+1}).
\label{dbls}
\end{equation}
Then~(\ref{sim}) shows that
$F_5\simeq Z(A^2G^3-A^3G^2)+Z(A^4G)-\zeta(5)$, thanks to the duality
$Z(A^2B^3)=Z(A^3B^2)$  of the MZVs with depth $d>1$.
Thus $Z(A^4G)$ cancels
in~(\ref{dbls}), to give $Z(A^2G^3-A^3G^2)\simeq\zeta(5)-\frac{17}{16}Z(A^4F)$,
and~(\ref{trins}) gives $A_5\simeq Z(8A^3G^2-8A^2G^3+6AG^4)$. Using~(\ref{sim})
to eliminate $Z(AG^4)\simeq\zeta(5)-Z(A^4F)$, we prove the claimed result~(\ref{a5}).
Then shuffle algebra restores the temporarily neglected products and lifts~(\ref{a5}) to give
\begin{eqnarray}
\sum_{n>0}\frac{(-1)^{n+1}}{n^5{2n\choose n}}&=&
\frac{5\,{\rm  Li}_5(\rho^2)-4\,\zeta(5)}{2}
-5\,{\rm  Li}_4(\rho^2)\log(\rho)
+4\,\zeta(3)(\log(\rho))^2\nonumber\\&&{}
+\left(\frac{2\pi}{3}\right)^2(\log(\rho))^3
-\frac43(\log(\rho))^5.
\label{a5full}
\end{eqnarray}

\subsection{Theorems at higher weight}

{\bf Theorem 3:} For weights $w<8$, every MZV and every MLV in the $\{A,F\}$
sub-alphabet is a ${\mathbb Q}$-linear combination of MGVs.

{\bf Proof:} For odd weights, the duality relation $Z(A^a B^b)=Z(A^b B^a)$ ensures
the cancellation of MZVs with depth $d>1$ in the reduction of~(\ref{dbls}) to MGVs and MZVs,
giving
\begin{equation}
\zeta(2n+1)\simeq4^n\sum_{k=2}^{2n}(-1)^kZ(A^{2n+1-k}G^{k})+Z(AG^{2n})
\label{zodd}
\end{equation}
from which  $Z(A^{2n}G)={\rm Li}_{2n+1}(\rho)$ is absent. All MZVs with $w<8$
are reducible to  $\zeta(2)$, $\zeta(3)$, $\zeta(5)$, $\zeta(7)$ and their products.
Thus~(\ref{zodd}) reduces them to MGVs and then Lemma~1 shows that
MLVs of the $\{A,F\}$ sub-alphabet reduce to MGVs, for $w<8$.~$\blacksquare$

{\bf Theorem 4:} There is at least one integer relation between
weight-6 MGVs and at least one between weight-8 MGVs.

{\bf Proof:} For even weight, $w=2n$, we obtain from~(\ref{dbls})
\begin{equation}
\sum_{k=2}^{2n-2}(-1)^kZ(A^{2n-k}G^{k})
\simeq 2Z(A^{2n-1}G)+(2^{1-2n}+1)Z(AG^{2n-1})+Z_{2n}
\label{red}
\end{equation}
where MZVs of depth $d>1$ contribute to the summands of
\begin{equation}
Z_w\equiv\sum_{k=2}^{w-2}(-1)^kZ(A^{w-k}B^{k}).
\label{z2n}
\end{equation}
However $Z_6\simeq0$, since there is no primitive MZV of weight 6,
and the MZV datamine~\cite{BBV} proves that $Z_8$ is
a rational multiple of $\pi^8$.~$\blacksquare$

In fact it is possible to prove that $Z_{2n}=4^{1-n}\zeta(2n)$, but that is not the issue.
The important point is that $Z_8$ does {\em not} contain the depth-2 primitive
$\zeta(5,3)\equiv{\rm Li}_{5,3}(1,1) $.
Hence we are stuck: we cannot extend Theorem~3 to higher weights
using the limited methods of this section.
Nor can we easily extend Theorem~4, since the argument by induction,
for lifting all neglected products to MGVs, cannot be relied on for $w>8$.

\subsection{Seed corn}

I have refrained from giving any experimental result in this section.
Hence we have learnt only a few things, thus far. The following are seminal.
\begin{enumerate}
\item MGVs of the golden sub-alphabet $\{A,G\}$ have at least one integer relation
at each of the even weights 4, 6 and 8.
\item Ap\'ery's alternating binomial sums reduce to ${\mathbb Z}$-linear sums of MGVs at all
weights $w>1$. For $w<6$ they have proven reductions to the depth-1 sums
${\rm Li}_w(\rho^p)$, with $p\in\{0,1,2\}$, and their products.
\item With the possible exception of $\zeta(5,3)$, we may reduce all MZVs with weight $w<9$ to MGVs.
\item We may reduce all MLVs of the sub-alphabet $\{A,F\}$ with $w<9$ to MGVs
and, if necessary, $\zeta(5,3)$.
\end{enumerate}

\section{A cornucopia of computation and conjecture}

Unless otherwise indicated, all further results are empirical
and all further claims are conjectural.
Hence I am now allowed to say that $\zeta(5,3)$ has an empirical
reduction to MGVs.

\subsection{The golden sub-alphabet for weights $w<9$}

Let $D_w$ be the number of  ${\mathbb Z}$-linearly independent MLVs of weight
$w$ in the full alphabet $\{A,B,F,G\}$. We shall  not be able to
determine $D_6$ until someone figures out how to prove the
irrationality of $(\zeta(3))^2/\pi^6$ (and much else).
However a plausible lower bound
is provided by numerical study of a sub-alphabet, using, for example,
the {\tt lindep} procedure of {\tt Pari-GP}~\cite{GP} to find integer relations
that are supported by many more digits of evidence than were required
to discover them. So the first question is clear: how many relations
are there between weight-5 Lyndon words in the golden sub-alphabet? I found precisely two
\begin{eqnarray}
10Z(AAGAG)&\simeq&Z(   8A^4G-138A^3G^2+111A^2G^3-114AG^4)\label{a51}\\
10Z(AGAGG)&\simeq&Z(-16A^4G  +8A^3G^2   -24A^2G^3    -3AG^4)\label{a52}
\end{eqnarray}
written here modulo 9 products of Lyndon words of lesser weight, which may
be restored by using the MLV datamine. These relations hold at 3000 digit
precision. Moreover, there is no further relation for $w<6$ with integer coefficients
of less than 200 digits.

I then found, with increasing labour, 4 relations between weight-6 Lyndon words,
modulo 19 products, and 8 relations between weight-7 Lyndon words, modulo 34 products.
Thus the number of relations forms the sequence $1,2,4,8\ldots$ starting at $w=4$.

Then came the {\em experimentum crucis}, at $w=8$, where I confidently
expected to find precisely 15 relations, modulo 66 products, for reasons that will emerge.
This required integer relation searches in a space of 82 dimensions,
each conducted at 3000-digit precision, in the hope that no integer coefficient
might have more than $\left\lfloor3000/82\right\rfloor=36$ digits, and was rewarded
by precisely 15 highly credible relations.

{\bf Conjecture 1:} The dimension of the space of ${\mathbb Z}$-linearly  independent
MGVs of weight $w$ is the tribonacci number $T_w $ defined by $T_1=1$, $T_2=2$,
$T_3=4$ and $T_w=T_{w-1}+T_{w-2}+T_{w-3}$ thereafter.

To see how this fits the data, consider the generating function
\begin{equation}
1+\sum_{w>0}T_w x^w=\frac{1}{1-x-x^2-x^3}\label{tgen}
\end{equation}
giving the tribonacci sequence, which for $w=1\ldots14$ is\\
{\tt 1, 2, 4, 7, 13, 24, 44, 81, 149, 274, 504, 927, 1705, 3136\dots}\\
Then the conjectured number $N_w$ of primitives forms the sequence\\
{\tt 1, 1, 2, 2, 4, 5, 10, 15, 26, 42, 74, 121, 212, 357\ldots}\\
for $w=1\ldots14$, determined by the filtration
\begin{equation}
\prod_{w>0}(1-x^w)^{N_w}=1-x-x^2-x^3.
\label{tprim}
\end{equation}
Now define a Lucas-type 3-step sequence, $S_w$, with
$S_1=1$, $S_2=3$, $S_3=7$ and $S_w=S_{w-1}+S_{w-2}+S_{w-3}$ thereafter.
Then a M\"obius transformation gives $w N_w=\sum_{d|w}\mu(w/d)S_d$.
The number $B_w$ of binary Lyndon words (discounting the illegal word $A$) grows
much faster, since  $w B_w=\sum_{d|w}\mu(w/d)(2^d-1)$. Hence the predicted number
of relations between Lyndon words is given by
\begin{equation}
R_w\equiv B_w-N_w=\frac{1}{w}\sum_{d|w}\mu(w/d)(2^d-1-S_d)
\label{trel}
\end{equation}
where the sum is over the divisors of $w$, $\mu(n)=0$ if $n$ is divisible by the
square of a prime and $\mu(n)=(-1)^{\omega(n)}$ when $n$ is a square-free integer
with $\omega(n)$ prime divisors. Then~(\ref{trel}) gives, for $w=4\ldots14$, the sequence\\
{\tt 1, 2, 4, 8, 15, 30, 57, 112, 214, 418, 804\ldots}\\
while the number of relations between MGVs, $2^{w-1}-T_w$,  gives\\
{\tt 1, 3, 8, 20, 47, 107, 238, 520, 1121, 2391, 5056\ldots}\\
which soon far exceeds the number $R_w$ of relations between Lyndon words.

Note that even at $w=5$ there is no
unique way of lifting the two relations~(\ref{a51},\ref{a52}) between Lyndon words
to the vector space of MGVs, where there are three relations. The third comes
from shuffling $Z(G)$ with relation~(\ref{id4}) at $w=4$. At $w=8$, there are
$2^7-T_8=128-81=47$ relations between MGVs of which only $R_8=15$
are irreducible relations between weight-8 Lyndon words. Hence the task for weight-8 MGVs
was to find 15 irreducible relations, each represented by a vector of $T_8+1=82$
integers, whose magnitude depends critically on how one has chosen to
eliminate MGVs of lower weight. We shall return to this  practical issue, in a later
subsection. For the present, it is sufficient to note that considerable
human intelligence may be needed to complete an investigation of the current scale;
if one is not careful, the problem may escalate, alarmingly, as occurred in
Erik Panzer's 73-dimensional integer-relation search,
needed to determine a 7-loop Feynman period in quantum field
theory~\cite{MDV,P}.

\subsection{The full $\{A,B,F,G\}$ alphabet of MLVs}

The harvest of 47 relations between 128 weight-8  MGVs
may seem rather modest. I now make the bold claim that these were
merely seed corn.

{\bf Conjecture 2:} The dimension of the space of ${\mathbb Z}$-linearly  independent
weight-$w$ MLVs in the alphabet $\{A,B,F,G\}$ is the tribonacci number $T_w$.

If this be true, then there is a glut of 36,783 integer relations between MLVs at $w=8$.
I claim to have recorded all of these in the MLV datamine.

\subsection{Rival bases for vector spaces}

{\bf Conjecture 3:} A basis for the vector space of MLVs of weight $w$
is provided by taking all the words of length $w$ in the
sub-alphabet $\{A,G\}$ that neither end in $A$ nor contain $A^3$.

This agrees with the MLV datamine. However,
the simplistic basis of Conjecture~3 was not the one used to compile the data
that  confirm the conjecture, for $w<9$, at overwhelming numerical precision.

The golden section $\rho=(\sqrt{5}-1)/2$ satisfies $1=\rho+\rho^2$.
Conjecture~3 singles out the letter $G=-d\log(1-\rho x)$, which seems
a little unfair on $F=-d\log(1-\rho^2x)$. Let's remedy that.

{\bf Conjecture 4:} A basis for the vector space of MLVs of weight $w$
is provided by taking all the words of length $w$ in the
sub-alphabet $\{A,F\}$ that neither end in $A$ nor contain $A^3$.

Again, this agrees with the datamine, which
can now act as an adjudicator between rivals,
whom I shall personify as $\rho$ and $\rho^2$.
Consider the determinant $|M|$ of the
$81\times81$ matrix $M$ that transforms the weight-8 MGVs of the $\{A,G\}$ basis of
Conjecture~3 to the weight-8 MLVs of the $\{A,F\}$ basis of Conjecture~4.
Every prime that occurs in the {\em denominator}  of $|M|$ is a bad mark for  $\rho$,
according to $\rho^2$, who correctly observes that $\rho$ must divide by that prime
when expressing MLVs in the $\rho$ basis.
But then, of course,  $\rho$ is equally correct in objecting that
every prime in the  {\em numerator} of $|M|$ is a bad mark for $\rho^2$,
who must divide by that prime to obtain MGVs.

Here are primes that are bad news for $\rho$:\\
{\tt 2, 7, 37, 41, 643, 198817, 2908441}\\
and here are primes that are bad news for $\rho^2$:\\
{\tt 3, 113, 1193, 1409, 7793, 2482024049916701}\\
from which it is clear that neither protagonist has cause to rejoice,
given that the datamine has no denominator prime greater than 11.

At this point, a third person, $-\rho$, speaks up on behalf of the letter
$H=-d\log(1+\rho x)$ claiming that the $\{A,H\}$ alphabet has been
unfairly excluded from this competition and pleading for it
to be enshrined in a further conjecture.
We may ignore this request, for the present, thanks to the following lemma.

{\bf Lemma 2:}  Let $f$ be a vector formed by the weight-$w$ MLVs defined in Conjecture~4
 and let  $h$ be the corresponding vector of iterated integrals  in which
 $F=-d\log(1-\rho^2x)$ is replaced by $H=-d\log(1+\rho x)$.
Then there is a unimodular integral matrix $U$ such that $h=Uf$.

{\bf Proof:} Let $W(A,B)$ be any binary word of length $w$
 that  neither ends in $A$ nor contains $A^3$.
Then $Z(W(A,H))$ is an element of $h$ and~(\ref{barr}) tells us that
 $Z(W( A,H))=Z(W(A+F,-F))$ is a sum of $2^{w-d}$ elements of $f$,
each with coefficient $(-1)^d$, where $d$ is the number of occurrences
of $B$ in $W(A,B)$.
Hence there is a matrix of integers $U_{i,j}$ such that $h=Uf$.
Now we elect to list the elements of vectors $f$ and $h$ by the
lexicographic order of $W(A,B)$. Then $U$ is upper triangular with
diagonal elements $\pm1$. Thus $|U|=\pm1$
and the transformation from $f$ to $h$ is unimodular, no matter how we
order the elements of $f$ and $h$.~$\blacksquare$

From this it is clear that every prime that is bad for $\rho^2$ is also bad
for $-\rho$ and that these primes are even worse than those
which are bad for $\rho$.

\subsection{Conjectural primitives}

Using a vector-space criterion for a putative basis is usually a bad idea,
for practical purposes.
It makes more sense to try to choose, at each weight $w>1$,
a set of primitives of the conjectured cardinality $N_w$ defined by~(\ref{tprim}),
which enumerates binary Lyndon words $W(A,B)$ that
do not contain $A^3$. It also enumerates
the Lyndon words that do not contain $B^3$. Moreover,
we may take the duals of those two choices, since the dual of a Lyndon word
is not necessarily a Lyndon word.
So now we have $2\times2=4$ choices for the set of words.

For each of those 4 choices of words,
we have 3 choices, $F$, $G$ and $H$, for what to substitute for the letter $B$,
which here stands merely as a binary place-holder.
Note that the choice of $H=-d\log(1+\rho x)$ is now not trivially
disposed of by Lemma~2, which was concerned
with a transformation between vector spaces,
not between primitives.
So we have (at least) $4\times3=12$ choices for sets of putative primitives.

{\bf Conjecture 5:} For each of these 12 choices, the specified words are primitive
and, with their products, provide a basis for all MLVs.

This agrees with the MLV datamine, for $w<9$.

\subsection{Depth-filtered primitives}

Being, by upbringing, an empiricist, I wished to reduce all 49,151 MLVs with $w<9$
to a set  of  primitives and  their products. That involved, {\em inter alia},
determination of 36,783 integer relations at $w=8$ in a search space of
dimension $T_8+1=82$, where even a good choice of primitives
might be expected  to yield integers with more than 15 digits. Inflation to
more than 35 digits, by products of denominator-primes,
such as $37\times41\times643\times198817\times2908441$,
could not be tolerated; some better idea was needed.

Having seen that the strong Conjecture~2, on
the number of primitives for the {\em full} alphabet, was viable at
low weights, I resolved not to restrict the primitives
to any sub-alphabet. Hence my Aufbau was based on ordering
primitive MLVs first by weight, $w$,  then by depth, $d$,
and finally, for each $w$ and $d$, by lexicographic order
in the $\{A,B,F,G\}$ alphabet.  Thus a systematic
choice of primitive words is given by
$$\{F\},\quad\{AB\},\quad\{A^2B,\,A^2G\},\quad\{A^3F,\,A^3G\},
\quad\{A^4B,\,A^4F,\,A^4G,\,A^3BF\}$$
for $w=1$ to 5, respectively.
Note that Landen's result of 1780, for the reduction of
${\rm Li}_3(\rho^2)-\frac45{\rm Li}_3(1)$, means that one must skip over
$A^2F$. The reducibility of $\zeta(4)$ requires the omission of $A^3B$.
At $w=5$, the first depth-2 word, $A^3BF$, is primitive. Then at $w=6$ the set
$$\{A^5F,\,A^5G,\,A^4FB,\,A^4GB,\,A^4GF\}$$
suffices, with the predicted cardinality $N_6=5$. At $w=7$, with $N_7=10$,
$$\{
A^6B,\,
A^6F,\,
A^6G,\,
A^5BF,\,
A^5BG,\,
A^5FB,\,
A^5GB,\,
A^5GF,\,
A^4FAB,\,
A^4FAG\}$$
likewise contains no depth-3 word.

So far, so good: no denominator-prime greater than 11 had appeared in any reduction
of a MLV with weight $w<8$. The products of these 25 primitives then supplied
$T_8-N_8=81-15=66$ elements of the conjectured vector space at $w=8$.
Then I worked my way through the 48 weight-8 MLVs that begin with $A^5$
and do not end in $A$, finding at high precision, that
\begin{eqnarray*}
&&\{
A^7F,\,
A^7G,\,
A^6FB,\,
A^6GB,\,
A^6GF,\,
A^5BAB,\,
A^5BAF,\,
A^5BAG,\nonumber\\&&
A^5FAB,\,
A^5FAG,\,
A^5GAB,\,
A^5GAF,\,
A^5BBF,\,
A^5BBG,\,
A^5BFG\}
\end{eqnarray*}
are independent primitives, with 3 of the 15 appearing at depth $d=3$.

This method is iterative: begin with a 67-dimensional integer relation search,
to determine the first primitive; when one is found
increment the partial basis; continue until a putative basis of dimension 81 is achieved;
the remaining checks are then in 82 dimensions.
With only 48 preliminary searches to perform, I could afford the luxury of
working at 3000 digits.

\subsection{Testing the conjectures at $w<9$}

Now comes the big question. Do {\em all} weight-8 MLVs reduce
to a 81-dimensional vector space?
The datamine, obtained by the judicious choice of the previous subsection,
attests that this is the case. It contains no denominator-prime
greater than 11.  All the MLVs were computed to 1250 digits but {\tt lindep}
was told only 1200 digits. It managed to find
credible 82-dimensional integer relations in all 36,783 cases with $w=8$. Then
I checked that these reproduce the extra 50 digits that
had been hidden from {\tt lindep}. So the probability of a spurious
reduction is less than $1/10^{45}$.

The largest prime found in a numerator of a coefficient of reduction was
158575062799, with 12 digits, in the coefficient of $\pi^8$ for the reduction of
$Z(AFGBGAFG)$. The largest integer found by {\tt lindep}
was $7\times15^3\times7908074791=186828266937375$, with 15 digits,
in the integer relation for $Z(GAFBGAFB)$. Had I used the basis
of Conjecture~3, the sizes of integers might have been inflated to 35 digits.

The datamine was compiled to test Conjecture~2.
Testing of the other conjectures, for $w<9$, is now simply a matter of matrix algebra.
Conjecture~1 asserts that the golden sub-alphabet, $\{A,G\}$,
gives vector spaces no smaller than those for the full alphabet.
To confirm this at weight $w<9$,  it suffices to find a set, with tribonacci cardinality,  $T_w$,
of weight-$w$ MGVs that are linearly independent, according to the rational vectors
of reduction in the datamine. This was easily done. Conjectures~3, 4 and 5
assert the validity of 14 concrete bases, confirmed
by the datamine, for $w<9$, by computing rational determinants
and showing that none vanishes. I have included these 14 naive choices in the hope that
someone might achieve, for one or more of them, a proof comparable to that
of Francis Brown, who showed~\cite{FBh} that Michael Hoffman's similarly naive
conjectural basis~\cite{MH} for MZVs is indeed functional at all weights,
though it lies beyond the wit of humankind to prove that there are no
further rational relations.

\section{Lewin's golden ladder combinations live for ever}

{\em \ldots glancing through the pages of Edwards' Calculus -- a fascinating book
if ever there was -- when my eye caught a paragraph at the foot of the page
recounting some formulae of Landen's}, wrote the electrical engineer,
Leonard Lewin, reminiscing~\cite{Lewin58}  about his schooldays in the 1930s.

I learnt about polylogarithms, more than 45 years ago,
from Lewin's book, {\em Dilogarithms and associated functions}, published~\cite{Lewin58} in 1958,
which also contained a good deal of information about polylogarithms and was much in demand
in my university's library,  by physicists calculating Feynman integrals. I studied it there,
again and again. Yet at that time, in the late 1960s, few mathematicians seemed to take interest
in this wonderful book.  It was something that harked back to seemingly miscellaneous results by
Euler, Landen~\cite{L6,L}, Spence, Abel, Hill~\cite{Hill1,Hill2}, Kummer, {\em et alia}. It was very useful for technical purposes,
like mine, but it seemed to be distinctly {\em pass\'e} in the upper echelons of courtiers of
the queen of the sciences -- pure mathematics -- as I perceived her, at that time.

A second edition, entitled
{\em Polylogarithms and associated functions}, was published~\cite{Lewin81} in 1981,
with little change of content, but now an encouraging preface by Alf van der Poorten.
Richard Askey remarked~\cite{Askey} that anyone who appreciates beautiful formulas should
become familiar with this book.

It was a source of satisfaction to me that modern mathematicians eventually
caught up with this fine scholar and engineer, and his enthusiastic readers
working in particle physics, by honouring Lewin with erudite contributions
to an American Mathematical Society  volume~\cite{Lewin91} entitled
{\em Structural properties of polylogarithms}, edited by him in 1991. Here he returned to Landen's
1780 formulas for ${\rm Li}_2(\rho)$, ${\rm Li}_2(\rho^2)$ and ${\rm Li}_3(\rho^2)$,
remarking that with $p\in\{1,2,3,4,6,8,10,12,20,24\}$ there are 6 combinations of
${\rm Li}_w(\rho^p)=\sum_{n>0}\rho^{pn}n^{-w}$
that are reducible to $\pi^2$ and $(\log(\rho))^2$ at $w=2$ and proceeding to find further
relations between
them, up to $w=9$.

\subsection{Six characters in search of an afterlife}

In addition to two dilogarithmic combinations from Landen~\cite{L} in 1780,
Lewin remarked on three from Coxeter~\cite{ Cox},
in 1935, and added a sixth finding~\cite{Lewin91} of his own.
The extension to weight $w$ is then as
follows. Let ${\cal S}_w$ be the set of the following 6 combinations of polylogarithms

\begin{eqnarray}
L_{1,w}&\equiv&{\rm Li}_w(\rho)\label{L1}\\
L_{2,w}&\equiv&{\rm Li}_w(\rho^2)\label{L2}\\
L_{6,w}&\equiv&{\rm Li}_w(\rho^6)
-2^w{\rm Li}_w(\rho^3)\label{L6}\\
L_{12,w}&\equiv&{\rm Li}_w(\rho^{12})
-2^{w-2}3\,{\rm Li}_w(\rho^6)
-3^{w-1}{\rm Li}_w(\rho^4)
\label{L12}\\
L_{20,w}&\equiv&{\rm Li}_w(\rho^{20})
-2^{w-1}{\rm Li}_w(\rho^{10})
-5^{w-1}3\,{\rm Li}_w(\rho^4)
\label{L20}\\
L_{24,w}&\equiv&{\rm Li}_w(\rho^{24})
+2^{w-1}{\rm Li}_w(\rho^{12})
-3^{w-1}2\,{\rm Li}_w(\rho^8)
-2^{2w-3}7\,{\rm Li}_w(\rho^6)\quad
\label{L24}
\end{eqnarray}
whose rule of construction is best appreciated at  $w=1$,  where
mere algebra proves
that $-2L_{n,1}/\log(\rho)=4, 2, -2, -1, -2, -1$,
for $n=1,2,6,12,20,24$, respectively. Thus, for example, the terms with $\rho$
raised to powers $p<24$ in the sixth case~(\ref{L24}), with $n=24$,
are included so as to exploit the identity
\begin{equation}
\rho^2(1-\rho^{24})^2(1-\rho^{12})^2=(1-\rho^8)^4(1-\rho^6)^7
\label{id24}
\end{equation}
which is an algebraic consequence of the defining property $\rho^2=1-\rho$
of the golden section. The generalization to $w>1$ is given
by the simple device of multiplying the required coefficient of ${\rm Li}_1(\rho^p)$ in
$L_{n,1}$ by $(n/p)^{w-1}$. Since all six elements of ${\cal S}_w$ have the required
property at $w=1$, so does any ${\mathbb Q}$-linear combination of them. I have chosen
to include as few terms as possible in the definitions, omitting divisors $p|n$ where $p$ is small
enough to be covered by a previous definition.

From this simple game, at $w=1$, we progress to a wonderful result for
the dilogarithms at $w=2$, where all 6 elements of ${\cal S}_2$ are rational combinations
of $\pi^2$ and $(\log(\rho))^2$. So it is natural to enquire what happens for $w>2$.

\subsection{Two insignificant departures from the ladder party}

Let $C_1=6$ and, for $w>1$, let $C_w$ be the number of
independent  ${\mathbb Z}$-linear combinations of elements of
${\cal S}_w$ that are reducible to $\zeta(w)$,
modulo products of polylogarithms of lesser weight.
Then we know that $C_2=6$, because
no dilogarithm leaves the ladder party.

At weight $w=3$, we know that $C_3<6$, because ${\rm Li}_3(\rho)\equiv Z(A^2G)$
and $\zeta(3)\equiv Z(A^2B)$ are independent primitive MLVs, or at least appear to be
so at 3000-digit precision. On the other hand, Landen proved that
${\rm Li}_3(\rho^2)\equiv Z(A^2F)\simeq\frac45Z(A^2B)$ remains at Lewin's ladder party.
Even better, there are $C_3=5$ independent combinations of trilogarithms in ${\cal S}_3$
that are reducible to $\zeta(3)$ and products.

At weight $w=4$, we know that $C_4<5$, because $Z(A^3F)$ and $Z(A^3G)$
are independent primitives of the $\{A,B,F,G\}$ alphabet and $Z(A^3B)$ is a multiple
of $\pi^4$ and hence not a primitive. Again it is notable that $C_4=4$
is as large as it could possibly be, given what we know about MLVs.

We may summarize the situation thus far by saying that for weights $w<5$
all 6 of the ladder combination evaluate as MLVs.  Those
polylogarithms that left the ladder were MLVs, by definition. Hence those that
remain must, despite their appearance, be combinations of MLVs, for $w<5$,
since Lewin found that they are linearly related to those that departed.

\subsection{A significant departure and arrival at weight 5}

By numerical methods, Lewin determined the sequence for $C_w$ as\\
{\tt 6, 6, 5, 4, 3, 2, 2, 1, 1, 0\ldots}\\
for $w=1\ldots10$, with no ladder relation remaining for weights $w>9$.
He charted the departures  in Figure~4.1 on page 52 of~\cite{Lewin91},
which shows, in the present notation, that $L_{6,5}$ leaves his party,
because no ${\mathbb Z}$-linear combination of
\begin{equation}
L_{6,5}\equiv{\rm Li}_5(\rho^6)-32\,{\rm Li}_5(\rho^3)=
-32\sum_{n>0}\frac{\rho^{6n-3}}{(2n-1)^5}
\label{L65sum}
\end{equation}
and the MLV polylogarithms ${\rm Li}_5(\rho^p)$, with $p\in\{0,1,2\}$, was found
by him to be reducible to products of polylogarithms of lesser weight.

This departee from Lewin's realm of ladder polylogarithms is greeted with joy
at the pearly gates of MLV-land, where $L_{6,5}$ is crowned in glory
as a primitive MLV of depth 2. The list of primitives systematically accumulated in Subsection~4.5
offers $L_{6,5}$ a home as proxy for $Z(A^3BF)={\rm Li}_{4,1}(1,\rho^2)$, since
numerical investigation quickly confirms that
\begin{equation}
L_{6,5}+\frac{165}{2}{\rm Li}_{4,1}(1,\rho^2)
\simeq165\,\zeta(5)-216\,{\rm Li_5}(\rho)+\frac{27}4\,{\rm Li_5}(\rho^2)
\label{L65ans}
\end{equation}
reduces to ${\rm Li}_5(\rho^p)$ with $p\in\{0,1,2\}$, modulo products of
terms of lesser weight that I here omit but are in the MLV datamine.
Metaphor apart, it is quite remarkable that, at the precise point where a depth-2 primitive MLV
is first required, the depth-1 sum in~(\ref{L65sum}) appears as its proxy. This discovery seemed to me to be
a fine compensation for the failure of MLVs to close under stuffles and led me to a wider conjecture.

\subsection{A conjecture for ladder combinations at all weights}

{\bf Conjecture 6:} For every weight $w$, the elements~(\ref{L1}) to (\ref{L24})
of ${\cal S}_w$ are ${\mathbb Q}$-linear
combinations of the tribonacci number $T_w$ of independent MGVs.

Combining Conjectures 2 and 6 with Lewin's list of departures, we predict
that the depth-1 sums $L_{6,6}$ and $L_{12,6}$ can stand as proxies for two
combinations of the weight-6 depth-2  primitive words $A^4FB$, $A^4GB$ and $A^4GF$.
Numerical computation confirms this
and also shows that $L_{12,6}$ is absent from $A_6$, whose primitive part
contains only {\em odd} powers of $\rho$:
\begin{equation}
A_6=Z((2A+G)^4G^2)\simeq\frac{32}{99}\sum_{n>0}
\frac{81\rho^{2n-1}-4\rho^{6n-3}}{(2n-1)^6}.
\label{a6}
\end{equation}

At $w=7$, there is no new arrival. As predicted, $L_{6,7}$ and $L_{12,7}$
stand as proxies for primitive MLVs with $w=7$ and $d=2$.

At $w=8$, Lewin observed another departure, leaving only one ladder combination.
Here the situation becomes rather interesting, since my method determined
that 3 of the 15 primitive weight-8 MLVs have depth $d=3$. The new arrival $L_{20,8}$
combines with $L_{12,8}$ and $L_{6,8}$, giving proxies for one primitive at $d=3$
and two at $d=2$.

Thus Conjecture~6 is neatly confirmed, at 3000-digit precision, for $w<9$, by reductions of all
6 of the ladder combinations, at each of those 8 weights, to a datamine basis of
tribonacci dimensionality $T_w$.

At $w=9$, there can be no new arrival, since Lewin observed no departure.

At $w=10$, the last departure occurs: no ladder combination survives. Thus Conjecture~6 predicts
that $L_{n,10}$, with $n\in\{6,12,20,24\}$, will serve as proxies for 4 weight-10 primitive MLVs
of depth $d>1$. I cannot determine how they distribute themselves by depth, since
$T_{10}=274$ is too large a basis size for me to handle, with current methods.

\section{Comparisons with roots of unity in physics}

The quantum field theory of the standard model of particle physics
leads to Feynman diagrams that define integrals whose
evaluation often yields multiple polylogarithms of the type
${\rm Li}_{a_1,a_2,\ldots,a_d}(z_1,z_2,\ldots,z_d)$ defined in~(\ref{Li}).

When there is a single large scale in the problem, set by a large
external energy or a large internal mass, the neglect of smaller
physical quantities, such as the masses of light quarks,
leads to arguments $z_j$ that are algebraic numbers.
These are often roots of unity~\cite{GBU}, with $z_j^N=1$ for
some modest value of $N$, with $N=1,2,6$ being prominent.
Hence a great deal  of effort has been expended on trying to understand
structural properties of iterated integrals defined by words
in alphabets with the letter $A=d\log(x)$ and other letters of the
form $-d\log(1-z_jx)$ with
$z_j^N=1$. These have a shuffle algebra, but not necessarily
a stuffle algebra, for which one needs to include
all of the $N$th roots of unity, while the physics may require only
a subset.

There has been fruitful
dialogue between highly focused physicists, who need to compute such numbers,
algebraic geometers, who are interested in  the parametric integrands that
produce them, and number theorists, who are interested in the
periods that result~\cite{BS}. Thus it is that I
have had the privilege of interaction with people like
Spencer Bloch, Francis Brown, Pierre Cartier, Alain Connes, Pierre Deligne,
Sasha Goncharov and  Don Zagier, who have patiently instructed me
in issues of importance to mathematicians and courteously
attended to accounts from the frontier of concrete
calculation that is an imperative for the standard model
of particle physics.

\subsection{The most puzzling root of unity}

From the perspective of iterated integrals,
the most puzzling root of unity is unity itself. The case $N=1$
gives MZVs and for these the infamous~\cite{FBbk}  Broadhurst-Kreimer (BK)
conjecture~\cite{BK} provides a bizarre answer to what seems
to be a very simple question: how many independently primitive MZVs
in the $\{A,B\}$ alphabet are there at a given weight and depth?
The conjectured answer, $N_{w,d}$,
is generated by
\begin{equation}
\prod_{w>2}\prod_{d>0}(1-x^w y^d)^{N_{w,d}}=1-\frac{x^3y}{1-x^2}
+\frac{x^{12}y^2(1-y^2)}{(1-x^4)(1-x^6)}
\label{bk}
\end{equation}
which Dirk Kreimer and I proposed, nearly 20 years ago,
and has been tested, with considerable ferocity,  by
numerical and exact~\cite{BBV} methods.

It is an intriguing fact that
the expansion of $x^{12}/((1-x^4)(1-x^6))$
enumerates cuspforms of the fundamental
modular group. Here I refer the reader to~\cite{FBbk,FBd3}
and wish only to observe that setting $y=1$ in~(\ref{bk}) we
obtain the claim that the vector-space dimensions are Padovan numbers,
generated by $1/(1-x^2-x^3)$, in accord with Hoffman's
conjectured~\cite{MH} vector-space basis of
finite MZVs with words that contain 
neither $A^3$  nor  $B^2$.
It has been proven~\cite{FBh} by Brown, with
inspired assistance from Zagier, that Hoffman's conjectural
basis is functional; no-one knows how to prove that
the Padovan upper bound is tight. Hoffman's basis for MZVs, like that
in Conjecture~3 for MLVs, is very inefficient, because of its large
denominator primes. The depth-filtered basis of
the MZV datamine~\cite{BBV} is far preferable.

Parallels between MZVs and MLVs in the $\{A,B,F,G\}$ alphabet are clear.
There is good support for my conjectures
that $1/(1-x-x^2-x^3)$ generates
the vector-space dimensions for MLVs and that primitives
are provided by Lyndon words in $\{A,G\}$ that do not contain $G^3$.
However I have no idea, in general, of how primitives may be filtered by
depth. Subsection~4.5 gives the data for $w<9$, where 3 depth-3 primitives
appear at $w=8$, of which one may be replaced by depth-1 sums
in a ladder combination.

\subsection{A well behaved root of unity}

Minus one is a very well behaved root of unity. The case $N=2$, for
the alphabet $\{A,B,C\}$ with $C=-d\log(1+x)$, gives
alternating sums and obviously includes MZVs. Alternating sums were
mastered, conjecturally at least, before the BK conjecture,
by my claim~\cite{B} that  the generating function
$1-xy/(1-x^2)$ tells us everything about the filtration
of primitives by both weight and depth. I proposed to Deligne
an empirically viable  set of primitives: Lyndon words in $\{A^2,C\}$, which
he later proved~\cite{D} valid at all weights.
My study of $\{A,B,C\}$ began by observing that the
Fibonacci numbers, generated by $1/(1-x-x^2)$,
fitted the data on dimensions.
It was a question from one my sons, Stephen,
then aged about 15, on how to obtain the Fibonacci numbers from
Pascal's triangle, that led from an enumeration
by weight to filtration by both depth and weight.

There are striking parallels with the results for MLVs, found here.
Instead of the Fibonacci sequence, the MLVs
appear to follow the tribonacci sequence,
generated by $1/(1-x-x^2-x^3)$. If we strike out $x$
we enumerate MZVs; if we strike out $x^3$,
we enumerate alternating sums.

Moreover, the MGVs of the $\{A,G\}$ sub-alphabet appear
to span the vector space of the full $\{A,B,F,G\}$ alphabet of MLVs,
so that we get 4 letters for the price of 2, which is more
economical than my old finding that
$\{A,C\}$ spans $\{A,B,C\}$,
which gave the less enticing offer of 3 for the price of 2.

\subsection{The primitive sixth root of unity}

Let $D=-\log(1-\lambda x)$, where $\lambda=(1+\sqrt{-3})/2$
is a primitive sixth root of unity. Then $1=\lambda+1/\lambda$ and setting $y=1/\lambda$
in~(\ref{tilde}) we obtain
\begin{equation}
Z(W)=\sum_{W=UV} L(\widetilde{U},\lambda)L(\overline{V},\lambda)=Z(\widetilde{W})
\label{dtilde}
\end{equation}
since $L(V,1/\lambda)=L(\overline{V},\lambda)$.
Recalling that  $\overline{V}(A,B)\equiv V(A+B,-B)$, we prove
that every MZV is a ${\mathbb Z}$-linear combination of iterated integrals in the $\{A,D\}$ alphabet.
With $W=AB$ we obtain $\zeta(2)=-3Z(DD)$ and with $W=AAB$ we have
two evaluations of $\zeta(3)=Z(A^2B)=Z(AB^2)$ and hence obtain an integer relation between
weight-3 words in $\{A,D\}$,  namely $Z((2A^2+AD+DA+11D^2)D)=0$,
with $\zeta(3)=3Z((A^2+5D^2)D)$. Note that~(\ref{dtilde}) reduces
$\zeta(5,3)$ to the $\{A,D\}$ alphabet, in contrast
with the impasse for $\{A,G\}$ at weight 8, which limited Theorem~3 in Section~3.

Jonathan Borwein, Joel Kamnitzer and I conjectured~\cite{BBK}  that the
Fibonacci numbers enumerate the vector-space
dimensions of the $\{A,D\}$ alphabet and
Deligne later proved~\cite{D} that these are upper bounds for the $\{A,B,D\}$ alphabet,
with functional primitives provided by Lyndon words in $\{A,D\}$
that do not contain $D^2$. A
better choice for weights $w<12$ is given in my datamine~\cite{MDV}
for the multiple Deligne values (MDVs) in $\{A,B,D\}$.

There is a significant parallel between MDVs, based on
$1=\lambda+1/\lambda$, and MLVs,  based on $1=\rho+\rho^2$. In neither
case does the stuffle algebra close. Yet in each we have a very simple
set of conjectural primitives: Deligne omits Lyndon words in $\{A,D\}$
that contain $D^2$; Conjecture~5 asserts, {\em inter alia},
that in $\{A,G\}$ we may omit Lyndon words that contain $G^3$.

In the Deligne case,
the generating function
$1-x^2y/(1-x)$ tells us everything about the filtration
of primitives by both weight and depth.
I am unable to go from $1/(1-x-x^2-x^3)$ to a simple
two-variable generating function that accounts for the data on depths in
Subsection~4.5, or for its pattern after modification by ladder combinations.

\subsection{Boring roots of unity}

The cases $N=3,4,8$ were also mastered by Deligne~\cite{D}.
Here there is little of interest, by way of structure.
The enumeration of dimensions is trivial: $2^w$
for $N=3,4$ and $3^w$ for $N=8$.  Primitives
are supplied by Lyndon words in $\{A,-d\log(1-\lambda^2x)\}$
for $N=3$, in $\{A,-d\log(1-ix)\}$ for $N=4$  and in
$\{A,-d\log(1-\sqrt{i}x),-d\log(1+\sqrt{i}x)\}$ for $N=8$,
with no known relations between such Lyndon words,
for a given value of $N$.
So far, particle physicists have had no need of these  lacklustre cases.

\subsection{The full 7-letter alphabet for sixth roots of unity}

Here  the full alphabet is $\{A,B,C,D,\overline{D},E,\overline{E}\}$,
where $E=-d\log(1-\lambda^2x)$ and bars denote complex conjugation.
Filtration of primitives by weight and depth is generated by
$1-xy-xy/(1-x)$ which at $y=1$ gives the vector-space  dimensions
as Fibonacci numbers with even indices: \\
{\tt 1, 3, 8, 21, 55, 144, 377, 987, 2584, 6765, 17711, 46368\ldots}\\
for $w=1\ldots12$.  These are generated by $1/(1-3x+x^2)$.
Thus at weight $w=6$ we have a Fibonacci dimension
$(\rho^{-12}-\rho^{12})/\sqrt{5}=144$ that is the same
as that for $w=11$ in the $\{A,B,D\}$ sub-alphabet of
MDVs. I  have given in~\cite{GBU} a conjecture for the primitives, using
a well-defined subset of Lyndon words in $\{A,E,C\}$, namely
those which do not contain $AC$.

The full alphabet
was studied in~\cite{sixth}, where it was needed for evaluating
Feynman diagrams dominated by a large mass.
The sub-alphabet $\{A,B\}$ is relevant to Feynman diagrams~\cite{pisa,BK}
dominated by a large external energy, as in electron-positron
collisions producing hadrons. $\{A,B,C\}$ is of the essence~\cite{B}
for the magnetic moment of the electron, where the same mass
appears externally and internally. In a study of a 7-loop
diagram~\cite{P} that contributes to the running of the self-coupling
of the Higgs boson, Panzer encountered the sub-alphabet
$\{A,D,\overline{E}\}$ but found an empirical reduction to $\{A,D\}$,
with large denominator primes, which I cleaned up in~\cite{MDV},
using primitives in $\{A,B,D\}$ that are more practical  than Deligne's.

\subsection{Other roots of unity}

Deligne was silent in~\cite{D} on the cases $N=5$ and $N=7$. I reported
in~\cite{GBU} conjectural enumerations of double sums
in those cases, but failed to arrive at overall conclusions.
The present study came from the fact that the golden
section $\rho=2\sin(\pi/10)$ interacts strongly with
the $N=5$ problem.
For $w<6$, {\tt lindep} gives  MLVs as ${\mathbb Q}$-linear combinations
of iterated integrals in the 6-letter alphabet formed by
$d\log(x)$ and $-d\log(1-\exp(2n\pi  i/5)x)$ for $n=0\ldots4$.
Yet I am loath to elevate this observation
to a conjecture for all weights.
Even if it were the case, it would not be of much use.
For example, we know that for $w<6$ there is only one primitive
MLV with depth $d>1$ to consider, namely $Z(A^3BF)\equiv{\rm Li}_{4,1}(1,\rho^2)$,
or its proxy, $L_{6,5}\equiv{\rm Li}_5(\rho^6)-32{\rm Li}_5(\rho^3)$.
I determined that the 5th-root alphabet gives
vector spaces with empirical dimensions
3, 8, 22, 61 and 168, for $w=1\ldots5$,
and that
$L_{6,5}$ reduces to 5th-root words at $w=5$,
where the MLV dimension $T_5=13$ is much smaller.
It is hard to see what
is gained by embedding a 13-dimensional problem in a 168-dimensional problem.

\section{Conclusions}

This investigation turned out better than I had dared to hope.
\begin{enumerate}
\item Empirically, MLVs enumerate very simply by weight, following the tribonacci sequence
generated by $1/(1-x-x^2-x^3)$. Striking out $x$, one has the Padovan enumeration of
that subset of MLVs that are MZVs, well known to physicists.
Striking out $x^3$ one gets the Fibonacci numbers
that appear in 3 other enumerations: for alternating sums, for MDVs, and (with even indices)
for the full alphabet of sixth roots of unity, all of which figure in physics applications.
\item A credible filtration of primitive MLVs by both depth and weight seems to be much harder to guess.
One has only to look at the BK conjecture~(\ref{bk}) for MZVs to imagine how long
it might take to accumulate sufficient MLV data  to tackle this problem empirically.
\item A datamine is available, on request to the author, with reductions of
all MLVs with weight $w<9$ to a systematically constructed basis that results
in 3,357,257 coefficients of rational reduction, without dividing by any prime greater than 11.
Then numerical  data for merely 40 primitives enable very fast evaluation of all
of these 49,151  MLVs to 20,000 digits.
A Hoffman-type basis that naively omits words containing $A^3$
would have introduced large denominator-primes.

\item Ap\'ery's sums, $A_w=\sum_{n>0}(-1)^{n+1}n^{-w}/{2n\choose n}$,
with $w>1$, have proven reductions to MLVs. For $w<6$ they reduce
to depth-1 sums and their products; $A_6$ is conjectured to do so, in~(\ref{a6}).
\item The available evidence agrees with the conjecture that Lewin's ladder combinations
remain MLVs for ever. When they leave his ladder, by failing to reduce to depth-1 MLVs,
they provide proxies for primitive MLVs of greater depth. This compensates for the failure
of the stuffle algebra to close in the case of MLVs.
\item It is not clear whether there are corresponding phenomena in other
algebraic number fields, such the Lehmer field~\cite{CLZ} that sets the record~\cite{BB}
for the longest known ladder, reaching up to $\zeta(17)$.
\item  It would be good to have a proof of the reduction~(\ref{L65ans}), which may be
accessible by combining stuffles that leave the MLV alphabet
with generalized doubling relations of the type used in~\cite{BBV}.

\end{enumerate}

{\bf Note added:} On reading this paper, Pierre Deligne sent me an
interesting letter~\cite{letter}, suggesting a possible geometric origin of relations
between MLVs, based on 5 points $\{\infty,\,0,\,1,\,\rho^{-1},\,\rho^{-2}\}\subset{\mathbb P}^1$.
His intuition leads to the symmetry group of a dodecahedron on a compactified
Euclidean plane.  To my untrained ear, this resonates with Coexter's Section 8, 
{\em Regular polyhedra inscribed in the hyperbolic absolute}, in the work~\cite{Cox},
which fuelled Lewin's interest~\cite{Lewin82} in dilogarithms ${\rm Li}_2(\rho^p)$,  with $p>2$,
and hence mine.

\newpage

 {\bf Acknowledgements:} I thank Johannes Bl\"umlein, for kindly providing works~\cite{Hill1,Hill2}
by Hill that sparked my historical interests, Freeman Dyson for encouraging my alphabetical endeavours,
Neil  Sloane, whose on-line encyclopaedia of integer sequences~\cite{OEIS} has so often guided me,
Jiangqiang Zhao for reminding me of my work in~\cite{BBK}, and the Erwin Schr\"odinger Institute
at the University of Vienna, for hospitality during a two-week meeting on the interrelation between mathematical physics,
number theory and non-commutative geometry, during which Erik Panzer and I
enjoyed lively dialogues on all manner of things related to~\cite{GBU}.

\raggedright

}\end{document}